\documentclass{jfm}

\usepackage[T1]{fontenc}
\usepackage[utf8]{inputenc}
\usepackage{newtxtext}
\usepackage{newtxmath}

\usepackage{amsmath}
\usepackage{amssymb}
\usepackage{graphicx}
\usepackage{epstopdf,epsfig}
\usepackage{subcaption}
\usepackage{xcolor}
\usepackage{natbib}
\usepackage{hyperref}
\hypersetup{
    colorlinks = true,
    urlcolor   = blue,
    citecolor  = black,
}

\makeatletter
\def\ps@titlepage{\leftskip\z@\let\@mkboth\@gobbletwo\vfuzz=5\p@
  \def\@oddhead{\hfil\llap{\thepage}}%
  \def\@evenhead{\rlap{\thepage}\hfil}%
  \def\@oddfoot{}%
  \def\@evenfoot{}%
  \def\sectionmark##1{}%
  \def\subsectionmark##1{}%
}
\def\@maketitle#1{%
 \newpage
 \vspace*{10\p@}\addvspace{5pc}%
 {\flushleft
  {\normalfont\LARGE\fontswitch\bfseries
  \put@rapidsHead%
  \@title@alignment%
  \@title \par}%
  \vskip 14\p@ \@plus 2\p@ \@minus 1\p@
  {\normalfont\large\fontswitch\bfseries\baselineskip=12\p@
     \@author@alignment%
     \lowercase{\@author}\par}%
  \vskip 4\p@ \@plus 1\p@
  {\normalfont\small
  \@aff@alignment%
  \@affiliation \par}%
  \vskip 8\p@ \@plus 2\p@ \@minus 1\p@
  \put@absrule
 \par}%
 \vskip 8\p@ \@plus 2\p@ \@minus 1\p@
}
\makeatother

\providecommand{\nek}{\textsc{Nek5000}}
\providecommand{\ellipsys}{\textsc{EllipSys}}

\providecommand{\redline}{\textcolor{red}{\rule[0.5ex]{1.6em}{0.45pt}}}
\providecommand{\blueline}{\textcolor{blue}{\rule[0.5ex]{1.6em}{0.45pt}}}
\providecommand{\dashedredline}{\textcolor{red}{\rule[0.5ex]{0.35em}{0.45pt}\hspace{0.16em}\rule[0.5ex]{0.35em}{0.45pt}\hspace{0.16em}\rule[0.5ex]{0.35em}{0.45pt}}}
\providecommand{\dashedblueline}{\textcolor{blue}{\rule[0.5ex]{0.35em}{0.45pt}\hspace{0.16em}\rule[0.5ex]{0.35em}{0.45pt}\hspace{0.16em}\rule[0.5ex]{0.35em}{0.45pt}}}

\title{Numerical simulations of transition and long-term response of a wind turbine airfoil}

\author{T. C. L. Fava\aff{1},
        N. S{\o}rensen\aff{2},
        D. S. Henningson\aff{1}
        \and A. Hanifi\aff{1}}

\affiliation{\aff{1}FLOW, Department of Engineering Mechanics, KTH Royal Institute of Technology, SE-100 44 Stockholm, Sweden
\aff{2}Department of Wind Energy, Technical University of Denmark, Ris{\o} Campus, Roskilde, Denmark}

\begin{document}

\maketitle

\begin{abstract}
Numerical simulations of a wind turbine airfoil of the FFA-W3 series corresponding to a section of the DTU 10-MW Reference Wind Turbine are performed. Wall-resolved large eddy simulations (LES) are carried out with the solvers \nek{} and \ellipsys{} for a chord Reynolds number of $1 \times 10^5$ and effective angle of attack $AoA=3.1^\circ-3.3^\circ$. It is shown that a domain width of $10\%$ of the chord is enough to capture the evolution of the main disturbances besides reproducing well the time-averaged flow. \ellipsys{} is validated against \nek{} for LES, indicating close results for the mean flow and most amplified perturbations. \ellipsys{} underpredicts the amplitude of Tollmien-Schlichting waves in the attached boundary layer due to a higher numerical dissipation but closely predicts the evolution of the Kelvin-Helmholtz (KH) mode in the laminar separation bubble. The latter is in close agreement with the predictions from parabolized stability equations (PSE). The shape of the mode is obtained with spectral proper orthogonal decomposition (SPOD), clearly showing the wavepacket of the KH mode forming in the LSB. The long-term evolution of the flow is computed with \ellipsys{}. A slow modulation of the normal force coefficient is identified with an amplitude of 10.5\% and a period of 48 flowthroughs, or equivalently a frequency $f=f^*c/U_\infty=0.021$ and Strouhal number $St=f \sin AoA=0.0012$.
	This frequency corresponds to low-frequency oscillations (LFOs) observed in several airfoil studies. However, $St$ is lower than previously noticed and occurs at a smaller $AoA$. In the DTU 10-MW Reference Wind Turbine, the period of these oscillations corresponds to 7.7 blade rotations. The periodic stalling and unstalling of the flow could trigger the LFO. The reverse flow on both sides of the airfoil is high enough to allow absolute instability, which may be responsible for the periodic bubble bursting.
\end{abstract}

\section{Introduction}

Flow separation often affects wind turbine blades due to elevated flow incidence angles and thick airfoils \citep{corten2001}. \citet{raju2008}  observed at least three main natural frequencies $f=f^* c / U_\infty$ ($f^*$, $U_\infty$, and $c$ are the dimensional temporal frequency, free-stream velocity, and chord length, respectively) in airfoils with separation due to wake shedding ($f=\mathcal{O}(1)$), Kelvin-Helmholtz modes  ($f=\mathcal{O}(10)$), and separation shedding ($f=\mathcal{O}(1)$). A separated shear layer over an airfoil is highly unstable, allowing the appearance of inviscid Kelvin-Helmholtz (KH) modes \citep{dovgal1994}. The central frequency of unstable KH modes typically determines the mean vortex shedding frequency $ f_s \theta_s/u_s=0.005-0.016$, where $\theta_s$ and $u_s$ are the momentum thickness and edge velocity at the separation point \citep{pauley1990,brinkerhoff2011}. 

The LSB may develop flapping, defined as the intermittent global motion of the separated shear layer upon separation \citep{jaroslawski2023}. This motion displays low frequencies, 100 times lower than the primary KH instability, i.e., $f=\mathcal{O}(10^{-1})$ \citep{zaman1989,rist2002}.
\citet{rist_vki} put forward three possible mechanisms for flapping. The first involves an LSB resonance to long-wavelength disturbances. The occurrence of absolute instability \citep{huerre1990}, which may cause such an oscillatory behavior, requires a minimum reverse flow of $12\%-25\%$ \citep{hammond1998,alam2000,rist2002,fasel2004,diwan2009,rodriguez2013} and the inflection point in the streamwise velocity profile to be located below the zero mass-flux line, i.e., $y_i<y_b$ \citep{avanci2019}.
The second considers a growth-decay mechanism \citep{cherry1984,dovgal1994}, consisting of several steps. Firstly, an incoming disturbance increase leads the LSB to shrink. The new, smaller LSB is less unstable, and disturbances are less amplified, allowing the LSB to enlarge over time. Eventually, the LSB grows big enough, shifting transition upstream and restarting the cycle. The third possibility is the interaction of the flow with the far-field boundaries or trailing edge noise \citep{arbey1983,nash1999,mcalpine1999,desquesnes2007,deng2007}. The frequency of the oscillations generated by resonance with the boundaries depends on the domain size, and the frequency of those generated by trailing-edge noise is of the order of the inverse of the flowthrough time. \citet{ehrenstein2008} found that the superposition of global modes, whose structures start inside the LSB and extend beyond the reattachment point, gives rise to an alternate cancellation yielding a low-frequency beating characteristic of flapping. The work of  \citet{cherubini2010}, in which the role of the non-normality of these modes in the occurrence of flapping was highlighted, agrees with those results. Furthermore, the experimental study by \citet{passagia2012} of the separated flow behind a bump identified the global modes predicted by \citet{ehrenstein2008}, whose structure presented a spanwise oscillation.

Low-frequency oscillations (LFO) characterized by strong lift variations have been identified experimentally in several studies \citep{jones1933,farren1935,armstrong1960,moss1979,zaman1987,zaman_potapczuk1989,bragg1993,heinrich1994,broeren2000,gurbacki2003,rinoie2004,tanaka2004,almutairi2013}. These oscillations typically have a Strouhal number in the range $St=(f^* c/U_\infty) \sin AoA=0.0048-0.03$ \citep{ansell2015}, where $AoA$ is the angle of attack.
The frequency of these oscillations is in the range of flapping, but the amplitude of the variations is much larger than that produced by the latter, suggesting another mechanism \citep{bragg1996}. In the experimental and numerical study of \citet{zaman1989}, the lift coefficient fluctuations reached $50\%$ of the mean value of this variable for an airfoil at $AoA=15^\circ$. Commensurate lift variations also occurred in the experimental work of \citet{bernardini2016}, considering a NACA 64$_3$-618 airfoil. Furthermore, \citet{zaman1989} found that these oscillations are hydrodynamic, with a phase speed $c_p=c_p^*/U_\infty=0.7$ and $St=0.02$. This frequency agrees with the results from \citet{farren1935} and \citet{zaman1987}, and it is one order lower than bluff-body shedding, which occurred for a deep-stall condition at $AoA=22.5^\circ$, with $c_p=0.95$ and $St=0.3$. The frequency $St=0.02$ is also one order lower than that of trailing-edge noise \citep{brooks1983,jones2010}. The phase speed of the low-frequency mode obtained by \citet{zaman1989} is much higher than that computed by \citet{ansell2015} ($c_p=0.145$) and \citet{crimi1974} ($c_p=0.25$), suggesting a large variability in the LFO modes. The low-frequency mechanism was attributed to the periodic stalling and unstalling of the flow on the suction side  \citep{zaman1989,broeren1998,sandham2008}. The maximum lift phase corresponds to a state with the longest LSB on the suction side. As time progresses, the pressure gradient flattens out, and the flow eventually becomes fully stalled. However, this reduces the adverse pressure gradient (APG), which allows the flow to reattach to form an LSB. The APG gradually increases again, and the reattachment point progressively moves downstream until the flow is totally separated again, restarting the cycle. Notice that the process where the LSB becomes a region of fully stalled flow is denominated bursting \citep{gaster1967}. Some works have associated bursting with the inception of absolute instability \citep{almutairi2013,negi2018a}. Nevertheless, LFO is essentially a two-dimensional phenomenon, unlike cases with three-dimensional stall cells \citep{broeren2001,sandham2008}. Notice that the described cycle can be captured with simple models, which contain the interaction between the boundary layer and potential flow \citep{sandham2008}. However, in order to capture correct frequencies, it seems that at least an LES approach is necessary \citep{almutairi2013}.  LFO has only been observed in airfoils with a thin-airfoil stall or a combination of thin-airfoil and trailing-edge stalls \citep{ansell2015}. Thin-airfoil stall happens when the flow separates due to an APG but reattaches owing to transition, forming a separation bubble \citep{almutairi2013}. Sufficiently high Reynolds numbers would act to reduce thin-airfoil stall and, consequently, LFO.
This phenomenon has been observed for $Re_c$ as high as $(1.0-5.8) \times 10^6$ \citep{mccullough1951,bragg1996,hristov2018,liu2020}. \citet{bernardini2016} showed that LFO may occur even in the post-stall regime and be triggered or suppressed with acoustic active control. 

More recently, \citet{tang2021} showed that merging shedding vortices could also generate a low-frequency oscillation of the order of previously observed LFO in airfoils. \citet{bouchard2022} found that the LFO on a stalled airfoil could be traced back to the upstream propagation inside the separated flow of relatively high-frequency ($St \approx 3$) perturbations generated at the trailing edge up to the leading edge, exciting low-frequency perturbations with $St=0.02$. \citet{aniffa2023} showed the occurrence of LFO on a flat plate with separated flow for high APG, where the shedding of vortices from the shear layer displayed an intermittent behavior. LFO was not significant for low APG. Further recent studies of LFO on an airfoil include \citet{eljack2020} and \citet{eljack2021}, which provided detailed accounts of the pre-stall and post-stall behaviors of this phenomenon. Recent wall-resolved simulations of rotating wind-turbine blade sections have further shown that rotation can modify the transition scenario by changing the amplification of Tollmien--Schlichting, Kelvin--Helmholtz and crossflow-type disturbances in the laminar separation bubble \citep{fava2024}. These results highlight the sensitivity of transition mechanisms on wind-turbine sections to rotation, Reynolds number and separation-bubble dynamics.

The current work studies transition and low-frequency oscillations on a thick wind turbine airfoil at $Re_c = 1 \times 10^5$ with wall-resolved (no subgrid-scale model) large eddy simulations (LES). For that, the incompressible Navier-Stokes solver \ellipsys, which is less computationally demanding than \nek, is validated against the latter and employed for the long-time integration of the equations. Linear stability analysis is used to assess the boundary layer stability. The paper is divided as follows: \S \ref{sec:methods} presents the airfoil geometry, modeling equations, flow solvers, and grid quality assessment. \S \ref{sec:width} shows the study of the dependence of the results on the domain spanwise width. \S \ref{sec:validation_ellipsys} exhibits the validation results of \ellipsys. \S \ref{sec:long_term} presents the study of the long-term variation of the lift coefficient. Finally, \S \ref{sec:conclusions} displays the conclusions of this work.

\section{Methods}\label{sec:methods}

\subsection{Problem modeling}

\begin{figure}
    \centering
    \begin{subfigure}[c]{0.5\textwidth}
        \includegraphics[angle=90,width=1\textwidth,trim={4cm 3cm 4cm 3cm},clip]{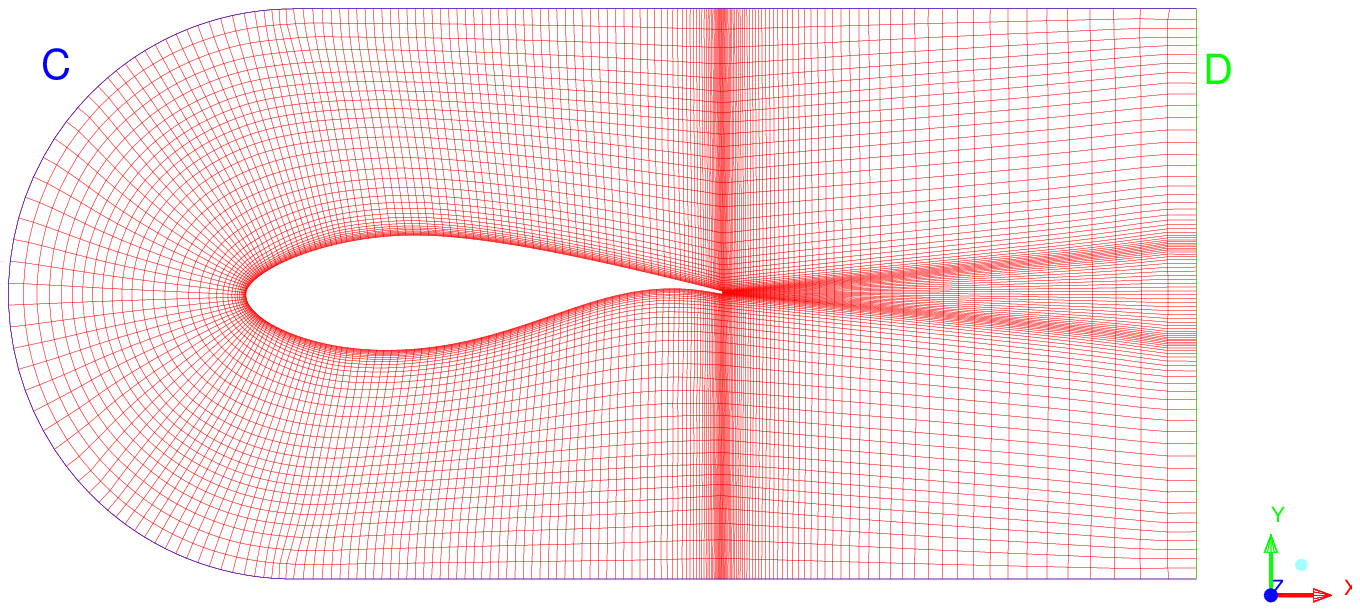}
    \end{subfigure}%
    \begin{subfigure}[c]{0.5\textwidth}
        \includegraphics[width=1\textwidth,trim={0cm 0cm 0cm 0cm},clip]{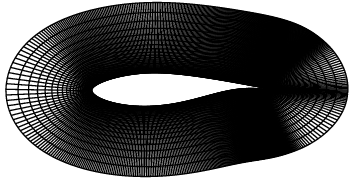}
    \end{subfigure}%
    
   \begin{subfigure}[c]{0.5\textwidth}
        \caption{\nek{}.}
        \label{fig:DNS_mesh}
    \end{subfigure}%
    \begin{subfigure}[c]{0.5\textwidth}
        \caption{\ellipsys{}.}
        \label{fig:mesh_EllipSys}
    \end{subfigure}%
    \caption{Computational domains and meshes employed in the simulations. Only the elements (without the Gauss-Lobatto-Legendre points) are shown in the \nek{} grid for enhanced visibility. In the \ellipsys{} grid, every third cell in each direction is exhibited.}
    \label{fig:domains}
\end{figure}

A blade section at $68\%$ of the radius of the DTU 10-MW Reference Wind Turbine \citep{bak2012} with a spanwise width of $10\%$ of the chord at a Reynolds number $Re_c = 1 \times 10^5$ is studied. The airfoil is a blend of $96\%$ of the FFA-W3-241 and $4\%$ of the FFA-W3-301 profiles \citep{bjorck1990}. The employed profile and computational domains in the simulations with the codes \nek{} and \ellipsys{} can be observed in Fig.~\ref{fig:domains}.

The equations modeling the problem are the incompressible Navier--Stokes equations given by

\begin{gather}\label{eq:governing_eq}
        \nabla \cdot \mathbf{u} = 0,\nonumber\\
        \frac{\partial \mathbf{u}}{\partial t} + \mathbf{u} \cdot \nabla \mathbf{u} = -\nabla p + \frac{1}{Re_c} \nabla^2 \mathbf{u},
\end{gather}

\noindent where $\mathbf{u}=(u_x,u_y,u_z)$, $p$ is the pressure, $Re_c=U_\infty c/\nu$ is the chord-based Reynolds number, $U_\infty$ is the relative free-stream inflow velocity, $c$ is the chord length, and $\nu$ is the kinematic viscosity. All quantities are non-dimensionalized by $U_\infty$ and $c$ if not otherwise stated. Although it is assumed that the flow arrives with an angle of attack $AoA=\arctan \left( \frac{V_\infty}{\Omega r} \right)=1.2^\circ$, where $V_\infty$ is the wind speed, $\Omega$ is the rotation rate, and $r$ is the radial location, rotation effects are not taken into account.
Notice that the relative velocity can also be written as $U_\infty = \sqrt{V_\infty^2+(\Omega r)^2}$.

As Dirichlet boundary condition for solving system \ref{eq:governing_eq}, $\mathbf{u}$ is specified at the inlet boundary, defined as surface C in Fig.~\ref{fig:DNS_mesh} (equivalently for \ellipsys{}). This boundary condition was obtained with Reynolds-Averaged Navier–Stokes (RANS) simulations carried out with \ellipsys{} (in RANS mode), whose results were interpolated on the boundaries of both \nek{} and \ellipsys{} computational domains. Moreover, a Neumann boundary condition of the form $\left[-p \mathbf{I}+1/Re_c\nabla \mathbf{u}\right] \cdot \mathbf{n}=0$, where $\mathbf{I}$ is the identity matrix and $\mathbf{n}$ is the outward-oriented normal unitary vector, is applied to the outlet plane (surface D in Fig.~\ref{fig:DNS_mesh}). Finally, periodicity boundary conditions are applied in the spanwise direction.

\subsection{Numerical methods}

The implicit large eddy simulations (LES) approach is taken, in which the incompressible Navier--Stokes equations (Eq. \ref{eq:governing_eq}) are solved without any turbulence model. Part of the simulations is performed with \nek{} \citep{fischer2008}, which is an open-source, highly scalable, and portable code based on the spectral element method (SEM), offering minimal dissipation and numerical noise, high accuracy, and nearly exponential convergence with the polynomial order $N$. SEM \citep{patera1984468} can be considered a high-order extension of the finite element method (FEM). SEM discretizes the computational domain into a finite number of non-overlapping elements. The basis functions, defined on each element, are Lagrange polynomials of order $N$. Here, $N=7$ was employed, providing converged results as shown in \citet{fava2023c}. The equations are solved in weak form using the $\mathbb{P}_N-  \mathbb{P}_{N-2}$ formulation, in which the velocity is expanded on Gauss--Lobatto--Legendre (GLL) points and the pressure on Gauss--Legendre (GL) points \citep{deville2002}. Third-order implicit backward differentiation (BDF), with an extrapolation scheme of order three for the convective term, is employed for time integration. The Courant number has been carefully assessed and is always below 0.3, corresponding to a fixed $\Delta t=5 \times 10^{-6}$ based on $U_\infty$ and $c$. Since SEM is characterized by very low numerical dissipation, which could lead to numerical instabilities, the highest $33\%$ of wavenumbers are filtered with an implicit filter that preserves the zero divergence of the flow \citep{negi2017}.
The filter amplitude is very low and corresponds to $2\times10^{-7}$ in an explicit filter.

The computations also employed \ellipsys{}, a general-purpose 3D solver developed at DTU/Risø \citep{Michelsen_1992,Michelsen_1994,Sorensen_1995}. The code solves the incompressible Navier--Stokes equations in curvilinear coordinates, using block-structured, finite volume discretization.
The formulation involves primitive variables (pressure-velocity) in a collocated grid. The SIMPLE algorithm is used to solve the pressure correction, with pressure (odd/even) decoupling being avoided using Rhie/Chow interpolation.
The convective terms are discretized with a second-order upwind scheme, whereas a central scheme is employed for the viscous ones. The equations are integrated in time using a second-order iterative time-stepping (dual time-stepping), where the equations are resolved iteratively with under-relaxation within each time step. \ellipsys{} is parallelized with MPI on distributed memory machines. Notice that \ellipsys{} has been widely used for RANS computations of transitional flows around rotors \citep{sorensen2009,sinem2020,fava2021}.
Nevertheless, the capabilities of \ellipsys{} for direct numerical or implicit large eddy simulations have not been explored.

\subsection{Spatial discretization}

\begin{figure}
    \centering
    \begin{subfigure}[t]{0.5\textwidth}
        \includegraphics[width=1\textwidth]
        {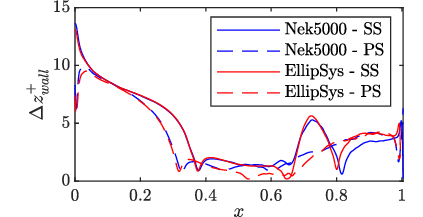}
        \caption{Spanwise resolution.}
        \label{fig:delta_z_plus_L1_o135n_bc0_EllipSys}
    \end{subfigure}%
    \begin{subfigure}[t]{0.5\textwidth}
        \includegraphics[width=1\linewidth]
        {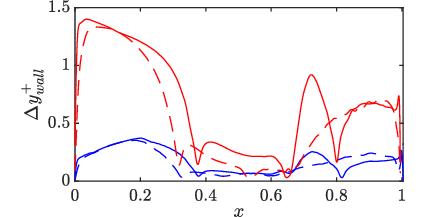}
        \caption{Wall-normal resolution.}
        \label{fig:delta_y_plus_L1_o135n_bc0_EllipSys}
    \end{subfigure}%

    \begin{subfigure}[t]{0.5\textwidth}
        \includegraphics[width=1\linewidth]
        {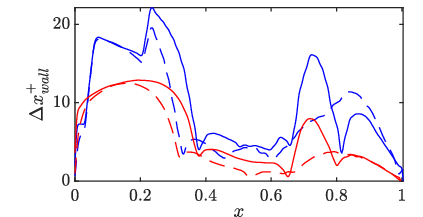}
        \caption{Streamwise resolution.}
        \label{fig:delta_x_plus_L1_o135n_bc0_EllipSys}
    \end{subfigure}%
    \caption{Wall grid resolution in the spanwise ($z$), wall-normal ($y$), and streamwise ($x$) directions on the suction side (SS, solid lines) and pressure side (PS, dashed lines).}
    \label{fig:delta_mesh}
\end{figure}

The \nek{} and \ellipsys{} computations are performed with grids with C and O topologies, respectively, as presented in Fig. \ref{fig:domains}. The number of grid points in the grids is summarized in Table~\ref{tab:number_points}. The points listed for the pressure and suction sides correspond to the streamwise direction.
In the wake, the points shown are in the streamwise direction in the C grid employed in \nek{} but in the wall-normal direction in \ellipsys{} due to the O topology. The grid resolution is further assessed by computing $\Delta \xi^+_{wall}=\Delta \xi_{wall}/(\nu \sqrt{\tau_w})$, where $\xi = x, y, z$. $\nu$ is the kinematic viscosity, and $\tau_w$ is the wall stress. $\Delta x^+_{wall}$ and $\Delta z^+_{wall}$ are taken as the maximum over each spectral element in the \nek{} computations, whereas $\Delta y^+_{wall}$ is relative to the height of the first GLL point. The results of this analysis are shown in Fig.~\ref{fig:delta_mesh}. The spanwise resolution is very similar in both meshes, as displayed in panel \ref{fig:delta_z_plus_L1_o135n_bc0_EllipSys}, and remains under 14. The wall-normal resolution in panel \ref{fig:delta_y_plus_L1_o135n_bc0_EllipSys} is higher in \nek{} ($\Delta y^+_{wall}<0.37$) because the GLL points tend to be clustered near the wall. If one considers the maximum over the element, the values get closer to that from \ellipsys{}. In the streamwise direction, displayed in panel \ref{fig:delta_x_plus_L1_o135n_bc0_EllipSys}, the resolution of \ellipsys{} is generally higher, but $\Delta x^+_{wall}<22$ in the \nek{} results. These values are close to that employed previously in high-fidelity implicit (no-model) LES of airfoils \citep{tanarro2020,atzori2020}, which proved to yield good agreement with DNS \citep{negi2018a}.

\begin{table}
  \begin{center}
  \begin{tabular}{lcccccc}
      Code          & Pressure  & Suction  & Trailing  & Wall-normal  & Wake & Span\\
                    & side      & side     & edge      & direction    &      &     \\[3pt]
      \ellipsys{}   & 969       & 962      & 120       & 100          & 100  & 160\\
      \nek          & 824       & 824      & 208       & 296          & 480  & 160
  \end{tabular}
  \caption{Number of grid points.}
  \label{tab:number_points}
  \end{center}
\end{table}

\section{Effect of the spanwise width of the domain}\label{sec:width}

The computational domain spanwise width $L_z$ is an essential parameter since it will determine what types of structures can be resolved. This is even more pressing in cases where there are laminar separation bubbles and detached flows \citep{almutairi2010,eljack2020} since the width of the structures forming compares with that of the streamwise dimensions of the LSB. The standard computational domain used by both \nek{} and \ellipsys{} presents $L_z=0.1$. The influence of this variable is studied by running a second simulation with \nek{} using $L_z=0.2$. This section presents the results of this analysis.

\subsection{Perturbation evolution}

\begin{figure}
    \centering
    \begin{subfigure}[t]{1\textwidth}
        \includegraphics[width=1\textwidth]
        {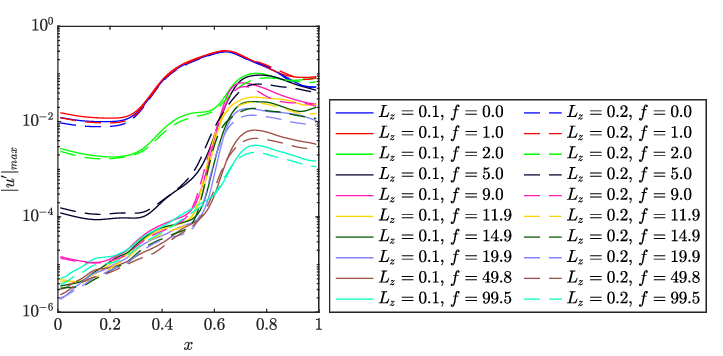}
        \caption{Maximum over $\beta$.}
        \label{fig:u_x_f_L1_o135n_bc0_L1_o135n_bc0_LD_SS_2}
    \end{subfigure}
    
    \begin{subfigure}[t]{1\textwidth}
        \includegraphics[width=1\linewidth]
        {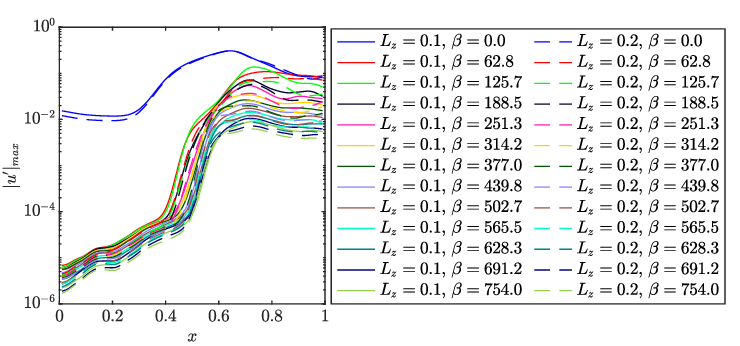}
        \caption{Maximum over $f$.}
        \label{fig:u_x_b_L1_o135n_bc0_L1_o135n_bc0_LD_SS_2}
    \end{subfigure}%
    \caption{Evolution of streamwise velocity fluctuations (maximum over the normal direction) for narrow ($L_z=0.1$) and wide ($L_z=0.2$) domains.}
    \label{fig:u_x_f_Lz02_Lz01}
\end{figure}

The streamwise evolution of streamwise velocity perturbations (maximum over the wall-normal direction and spanwise wavenumber $\beta$) is studied for selected frequencies, and the results are presented in Fig.~\ref{fig:u_x_f_L1_o135n_bc0_L1_o135n_bc0_LD_SS_2}. Steady ($f=0$) and low-frequency ($f=1$) perturbations display the highest amplitude. The curves for $L_z=0.1$ and $L_z=0.2$ are in close agreement. Furthermore, the agreement improved with the averaging time. Higher-frequency disturbances also display a good matching. Minor disagreements are present in the turbulent flow region towards the end of the domain, but this is expected as very long time series are needed to have converged turbulence statistics. Figure~\ref{fig:u_x_b_L1_o135n_bc0_L1_o135n_bc0_LD_SS_2} considers the maximum perturbation amplitude over the normal direction and frequency for given $\beta$. Two-dimensional ($\beta=0$) fluctuations present the highest amplitude, with much higher values than three-dimensional disturbances, until the flow becomes turbulent. There is good agreement between the $L_z=0.2$ and $L_z=0.1$ curves, although the amplitude for the narrower domain is generally slightly higher. This may be because intermediate spanwise wavenumbers are unresolved in the narrow domain, such as $\beta=2\pi(2n-1)/0.2$, with $n=1,2,\cdots$.
Therefore, energy in these modes must be redistributed to the resolved $\beta$ in the $L_z=0.1$ simulation, increasing their amplitude. Notice that $\beta=125.7$ (second spanwise harmonic) presents high amplitude, which even exceeds that of the first harmonic before transition. These modes are related to the spanwise deformation of the flow with the appearance of two spanwise wavelengths in the separation region.

\subsection{Instantaneous structures}

\begin{figure}
    \centering
    \begin{subfigure}[t]{0.49\textwidth}
		\centering
		\includegraphics[width=1\textwidth,trim={1.5cm 10cm 4cm 7cm},clip]{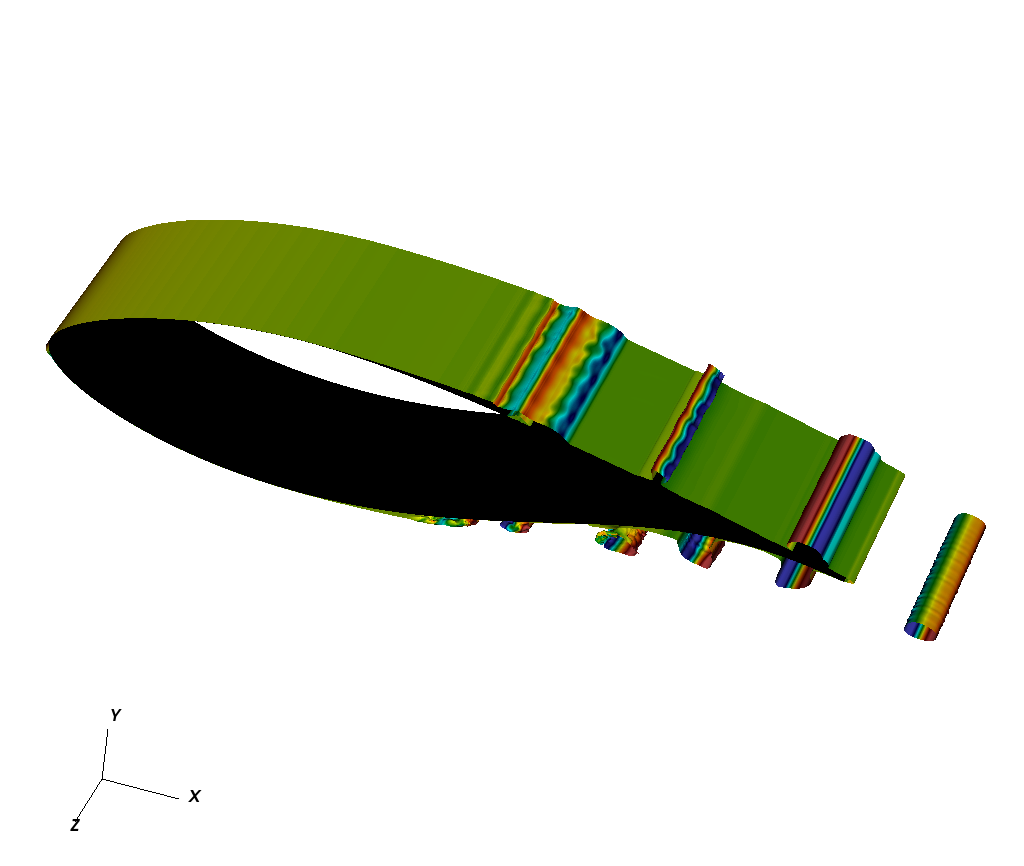}
        \caption{$u=0.1$ colored by $v$ for $L_z=0.2$.}
        \label{fig:L1_o135n_bc0_LD_3_2}
	\end{subfigure}    
    \begin{subfigure}[t]{0.49\textwidth}
        \centering
        \includegraphics[width=1\linewidth,trim={1.5cm 10cm 4cm 7cm},clip]
        {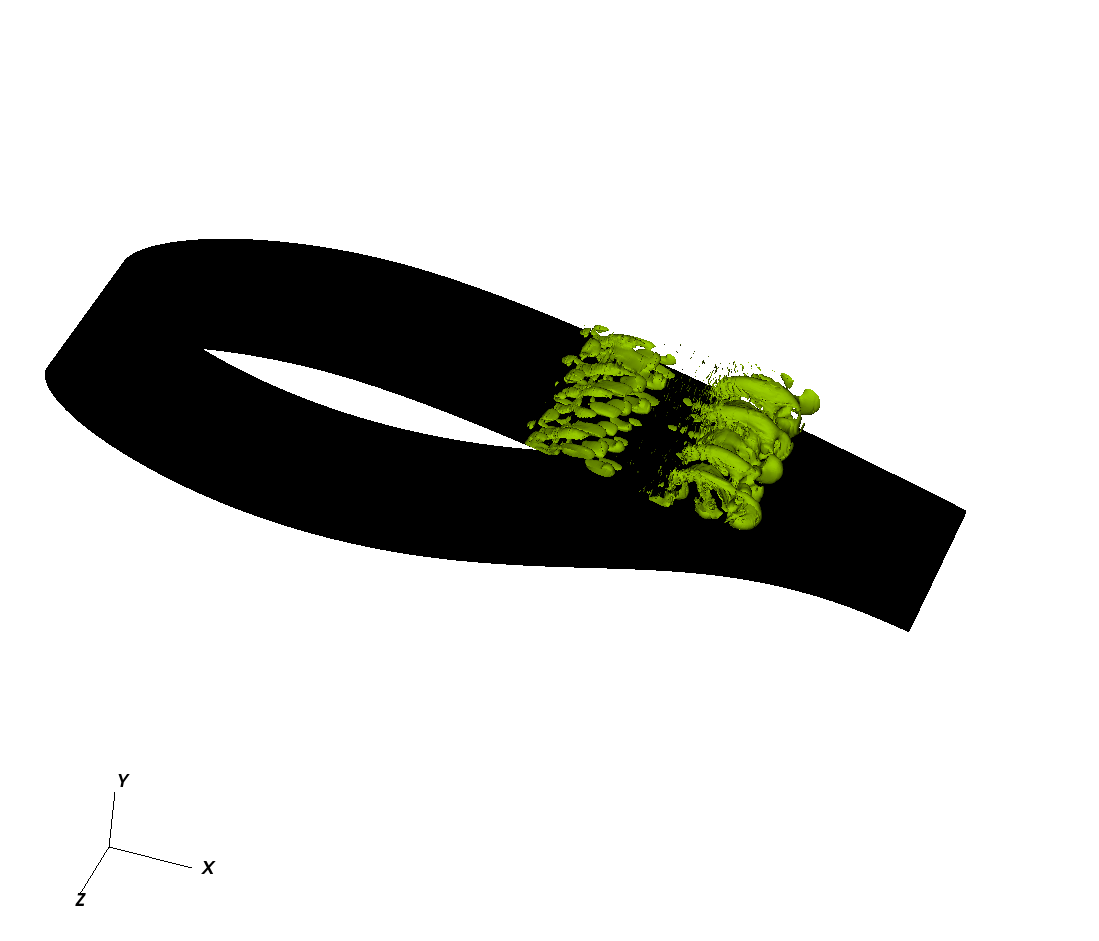}
        \caption{$w=0.01$ for $L_z=0.2$.}
        \label{fig:L1_o135n_bc0_LD_W_structures_3}
    \end{subfigure}
    
   \begin{subfigure}[t]{0.49\textwidth}
		\centering
		\includegraphics[width=1\textwidth,trim={1cm 9cm 4cm 9cm},clip]{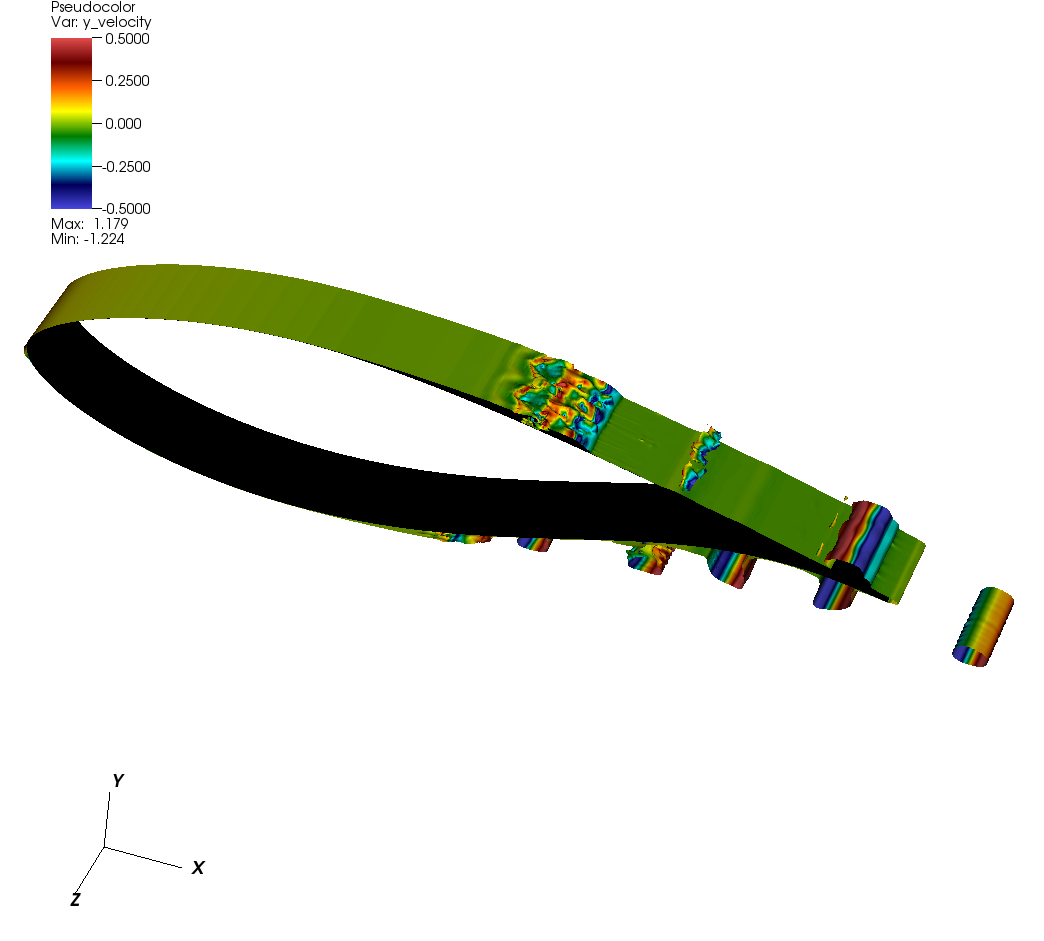}
        \caption{$u=0.1$ colored by $v$ for $L_z=0.1$.}
        \label{fig:L1_o135n_bc0_1}
	\end{subfigure}%
    \begin{subfigure}[t]{0.49\textwidth}
	    \centering
		\includegraphics[width=1\textwidth,trim={1cm 9cm 4cm 9cm},clip]{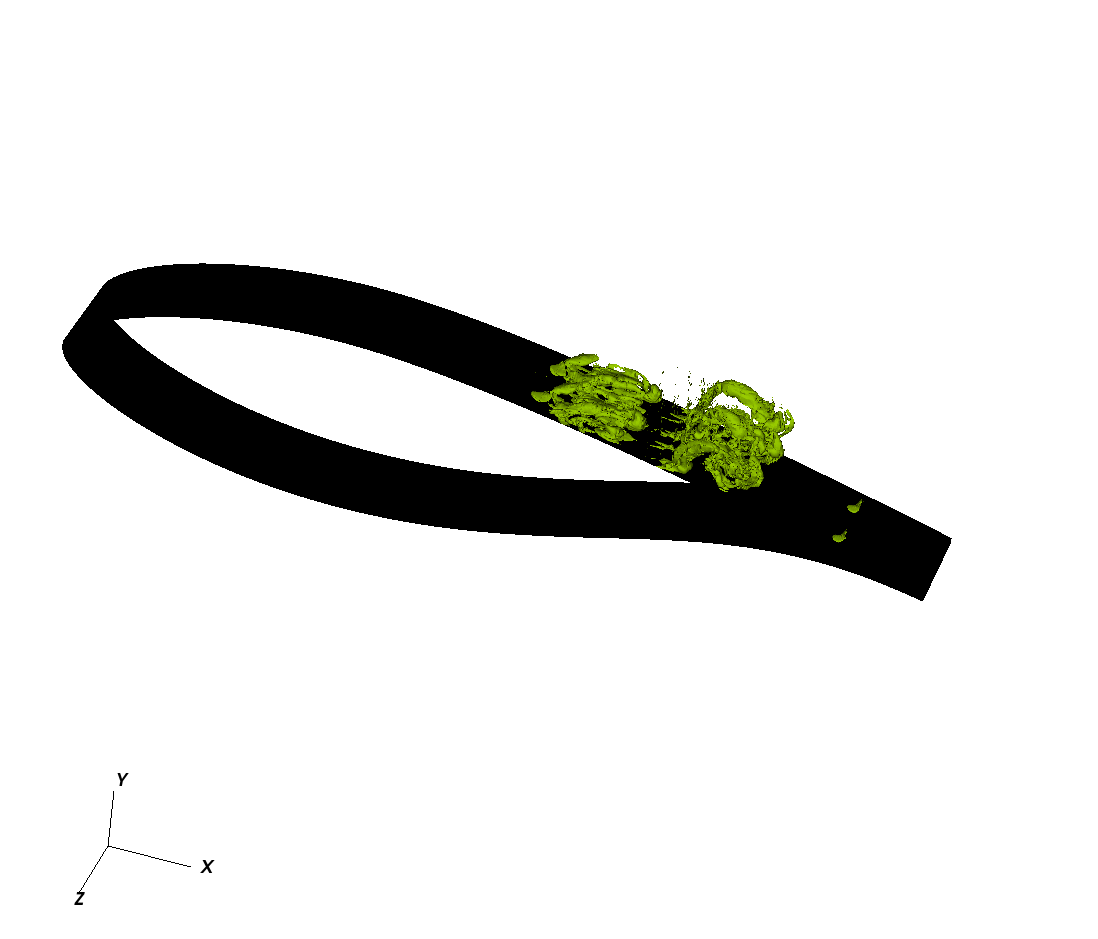}
        \caption{$w=0.05$ for $L_z=0.1$.}
        \label{fig:L1_o135n_bc0_W_structures_1}
    \end{subfigure}    
    \caption{Instantaneous isosurfaces on the suction side of the airfoil for narrow and wide domains at $T=2.0 c/U_\infty$.}
    \label{fig:structures_f_Lz02_Lz01}
\end{figure}

The instantaneous structures on the suction side of the airfoil for $L_z=0.1$ and $L_z=0.2$ are presented in Fig.~\ref{fig:structures_f_Lz02_Lz01}. In panels \ref{fig:L1_o135n_bc0_LD_3_2} and \ref{fig:L1_o135n_bc0_1}, two-dimensional rolls are formed over the laminar separation bubble. These structures correspond to Kelvin-Helmholtz (KH) vortices generated by the roll-up of the shear layer \citep{dovgal1994}. A spanwise modulation of these structures is visible in both cases, more evident in the narrow domain. The analysis of the isosurfaces of $w$ in panels \ref{fig:L1_o135n_bc0_LD_W_structures_3} and \ref{fig:L1_o135n_bc0_W_structures_1} indicate similar structures for the two domain widths, i.e., in the wider domain, the structure of the $L_z=0.1$ case appears repeated once more. This further proves that the narrow domain may be sufficient for the computations. Notice that there are two types of structures in these figures. The most upstream one resembles $\Lambda$ vortices, associated with the secondary instability of Tollmien-Schlichting (TS) waves \citep{klebanoff1962,herbert1988}. This instability is present in the front part of the LSB since the KH mode only becomes dominant further downstream, close to the LSB maximum height \citep{rist2002,diwan2009,jaroslawski2023}. Those further downstream structures are farther away from the wall and resemble the three-dimensional breakdown of a shear layer \citep{rogers1992}. Furthermore, \citet{he2017} found that both short- and long-wavelength (SW and LW) instabilities can occur in these types of transition. Despite the SW instability presenting the highest growth rate, the mode in the current simulations, with only two wavelengths in the spanwise direction for $L_z=0.1$, is characteristic of an LW mode.

\subsection{Time-averaged reverse flow}

\begin{figure}
    \centering
    \begin{subfigure}[t]{0.5\textwidth}
        \includegraphics[width=1\textwidth]
        {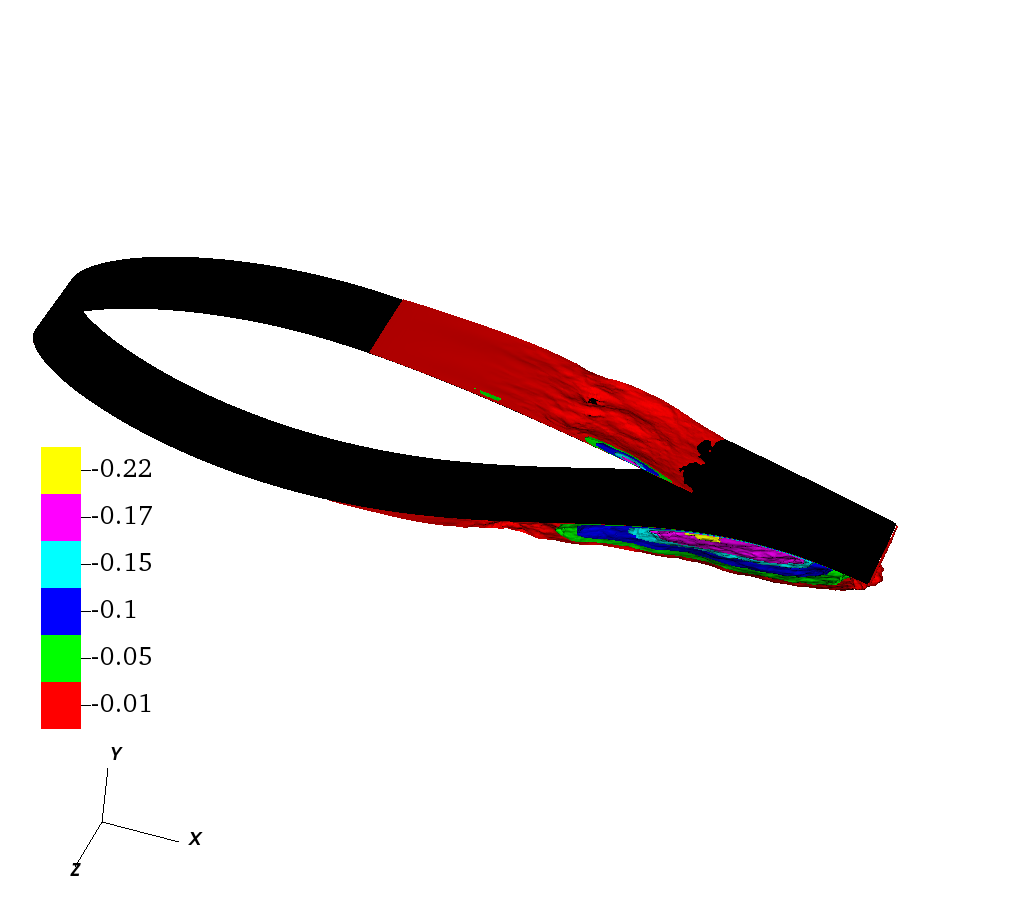}
        \caption{$L_z=0.1$.}
        \label{fig:mean_3D_L1_o135n_bc0_Tavg12}
    \end{subfigure}%
    \begin{subfigure}[t]{0.5\textwidth}
        \includegraphics[width=1\linewidth]
        {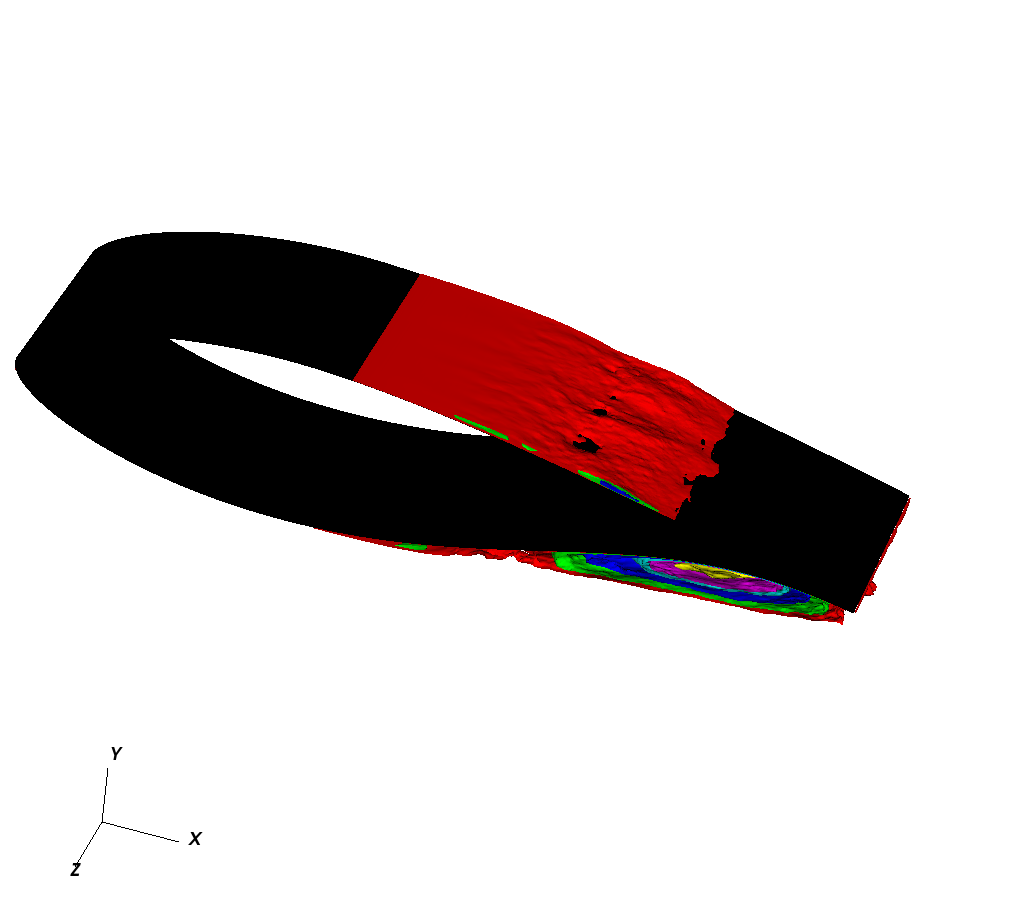}
        \caption{$L_z=0.2$.}
        \label{fig:mean_3D_L1_o135n_bc0_LD_Tavg6.png}
    \end{subfigure}%
    \caption{Isosurfaces of time-averaged streamwise velocity for several levels of reverse flow.}
    \label{fig:mean_3D}
\end{figure}

Figure~\ref{fig:mean_3D} presents the isosurfaces of time-averaged streamwise velocity for several levels of reverse flow. On the suction side, the flow presents a laminar separation bubble whose initial region is two-dimensional but develops a spanwise modulation further downstream, where the reverse flow is higher. This oscillation is characterized by two spanwise wavelengths in the narrow domain and double in the wide domain. The reattachment line presents a C shape, characteristic of the appearance of a steady global mode that leads to the three-dimensionalization of the reverse flow region \citep{rodriguez2009}. Such a mode typically requires a minimum reverse flow of -7\% \citep{rodriguez2010,rodriguez2013}, and the maximum reverse flow on the suction side is much stronger than that, reaching -17\%. The flow displays a trailing-edge stall on the pressure side \citep{mccullough1951}, where the maximum reverse flow is -22\%. These values of reverse flow are potentially enough to trigger absolute instability on both sides of the airfoil \citep{huerre1990}.

\section{Validation of \ellipsys{} for wall-resolved large eddy simulations}\label{sec:validation_ellipsys}

Once the narrow domain is shown to be enough for correctly capturing the flow physics, one can proceed with the validation of \ellipsys{} for wall-resolved large eddy simulations (LES), i.e., LES without subgrid-scale model. Validating \ellipsys{} is desirable because of the possibility of running cases with a larger domain and longer time. This is enabled by the ability of \ellipsys{} to deal with non-conformal meshes (now also available in the adaptive mesh refinement implementation of \nek{} \citep{offermans2019}), and a lower order of error truncation, which may be enough for the desired studies of transition.

\subsection{Mean flow fields}

\begin{figure}
    \centering
    \begin{subfigure}[t]{0.5\textwidth}
        \includegraphics[width=1\textwidth]
        {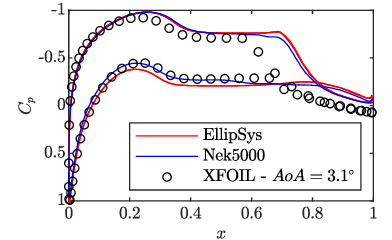}
        \caption{Pressure coefficient.}
        \label{fig:Cp_L1_o135n_bc0_EllipSys_2}
    \end{subfigure}%
    \begin{subfigure}[t]{0.5\textwidth}
        \includegraphics[width=1\linewidth]
        {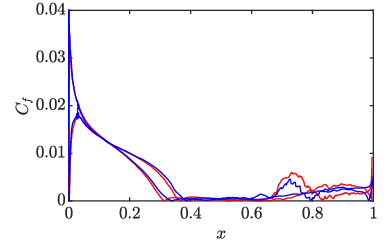}
        \caption{Wall friction coefficient.}
        \label{fig:Cf_L1_o135n_bc0_EllipSys_2}
    \end{subfigure}%
    \caption{Comparison of pressure and wall-friction coefficients.}
    \label{fig:Cp}
\end{figure}

\begin{figure}
    \centering
    \includegraphics[width=0.5\textwidth]
        {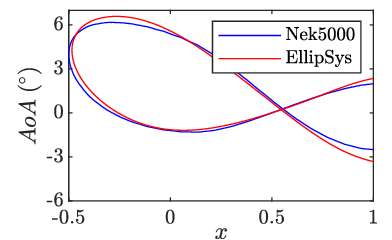}
    \caption{Comparison of the incidence angle on the domain boundaries.}
    \label{fig:AoA_L1_o135n_bc0_EllipSys}
\end{figure}

The pressure coefficient ($C_p$) obtained with the two solvers is compared in Fig.~\ref{fig:Cp}. In general, a good agreement between \ellipsys{} and \nek{} results is obtained, especially on the suction side of the airfoil (top part of the curves). Certain discrepancies exist between the curves in $x=0.15-0.55$ on the pressure side where \nek{} predicts a lower pressure. There are two primary sources for the differences between the codes. The first is the averaging time that is reduced in \nek{} since the computational cost is much higher than that for running \ellipsys{}. The second is the slightly smaller computational domain in \ellipsys{}, which may constrain the induced velocity, reducing the downwash and leading to a slightly higher effective angle of attack. The variation in the incidence angle on the domain boundaries is compared in Fig.~\ref{fig:AoA_L1_o135n_bc0_EllipSys}, where it is clear that the $AoA$ in \ellipsys{} is slightly higher. This difference is low on the leading-edge plane, in the range $0.06^\circ-0.2^\circ$ or $3.7\%-5.3\%$. The effective angle of attack in the simulations, which is typically different from the boundary conditions, is estimated by finding the $AoA$ for which XFOIL \citep{Drela_1989} predicts the closest pressure distribution to that from \nek{} and \ellipsys{}. An effective $AoA=3.1^\circ$ is obtained from this analysis for \nek{}, as shown in Fig.~\ref{fig:Cp}. The corresponding value for \ellipsys{} is slightly higher, $AoA=3.3^\circ$. Notice that XFOIL mispredicts $C_p$ in the region with thick separated shear layers. Regarding the wall-friction coefficient ($C_f$) in Fig.~\ref{fig:Cf_L1_o135n_bc0_EllipSys_2}, there is good agreement between \nek{} and \ellipsys{} until transition occurs, characterized by the increase in $C_f$ at $x=0.65$.

\begin{figure}
    \centering
    \begin{subfigure}[t]{0.5\textwidth}
        \includegraphics[width=1\textwidth]
        {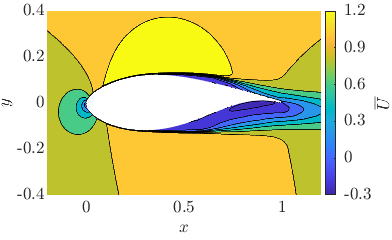}
        \caption{$x$ velocity from \ellipsys{}.}
        \label{fig:Ucontours_EllipSys_2}
    \end{subfigure}%
    \begin{subfigure}[t]{0.5\textwidth}
        \includegraphics[width=1\linewidth]
        {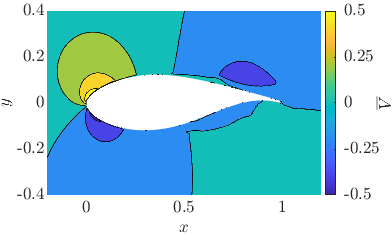}
        \caption{$y$ velocity from \ellipsys{}.}
        \label{fig:Vcontours_EllipSys_2}
    \end{subfigure}%

    \begin{subfigure}[t]{0.5\textwidth}
        \includegraphics[width=1\textwidth]
        {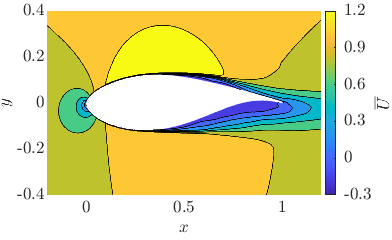}
        \caption{$x$ velocity from \nek{}.}
        \label{fig:Ucontours_L1_o135n_bc0_2}
    \end{subfigure}%
    \begin{subfigure}[t]{0.5\textwidth}
        \includegraphics[width=1\linewidth]
        {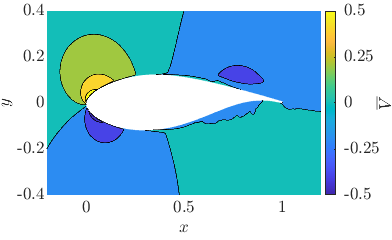}
        \caption{$y$ velocity from \nek{}.}
        \label{fig:Vcontours_L1_o135n_bc0_2}
    \end{subfigure}%
    \caption{Comparison of the spanwise- and time-averaged horizontal ($\overline{U}$) and vertical ($\overline{V}$) velocities in the near field of the airfoil.}
    \label{fig:U_contours}
\end{figure}

The spanwise- and time-averaged flow around the airfoil is depicted in Fig.~\ref{fig:U_contours} for the horizontal ($\overline{U}$) and vertical ($\overline{V}$) velocities. Overall, good agreement is obtained for both components, especially in the front part of the airfoil. Discrepancies occur in the laminar separation bubble (LSB) on the suction side and trailing-edge separation region on the pressure side, where \ellipsys{} predicts a higher reverse flow and thicker shear layer. The over-speed region on the suction side is slightly smaller than that in \nek{} due to the lower effective $AoA$.

\begin{figure}
	\centering
	\includegraphics[width=\linewidth]{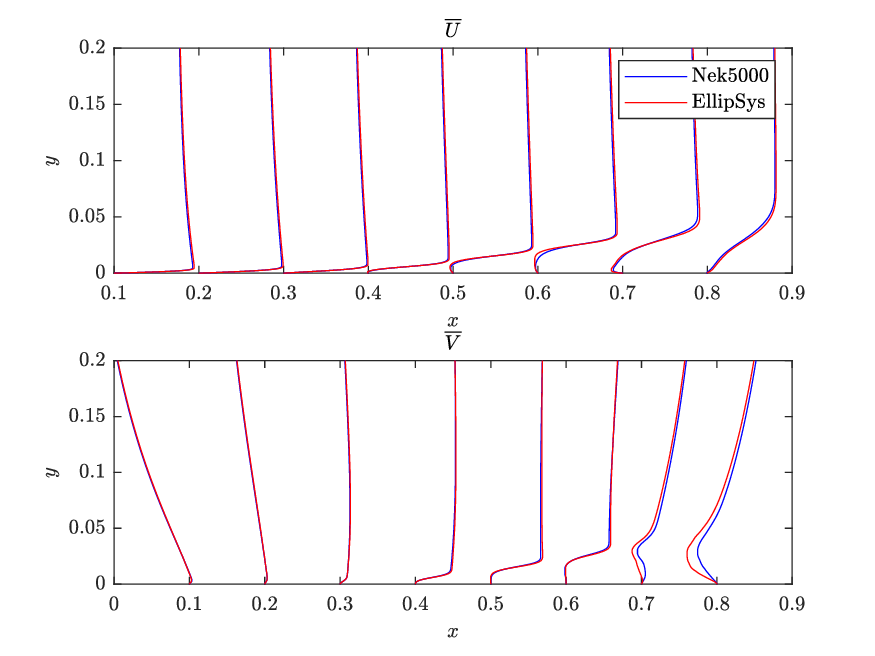}
	\vspace{-1cm}
	\caption{Comparison of the normal profiles of streamwise and normal velocities on the suction side. A scale of 0.1 corresponds to a velocity of 1.4 in the plot of $\overline{U}$ and 0.2 in the one of $\overline{V}$.}.
    \label{fig:vel_prof_SS_comparison}
\end{figure}

A more accurate comparison between the near-wall velocity on the suction side is carried out in Fig.~\ref{fig:vel_prof_SS_comparison}, where the spanwise- and time-averaged wall-normal profiles of streamwise ($\overline{U}$) and wall-normal ($\overline{V}$) velocities from \nek{} and \ellipsys{} are displayed. Close agreement is obtained for $\overline{U}$, where only minor differences are observed in the maximum reverse flow at $x=0.6$. The velocity profiles of $\overline{V}$ are also in close agreement for $x \le 0.6$, with differences only apparent in the reattached turbulent boundary layer. The agreement improved with the averaging time, indicating that a longer collection of statistics would further reduce the discrepancies. As expected, the results on the pressure side display large differences in the separated shear layer region, which, as discussed above, may be more affected by the slightly lower effective angle of attack in \nek{}.

\begin{figure}
	\centering
	\includegraphics[width=\linewidth]{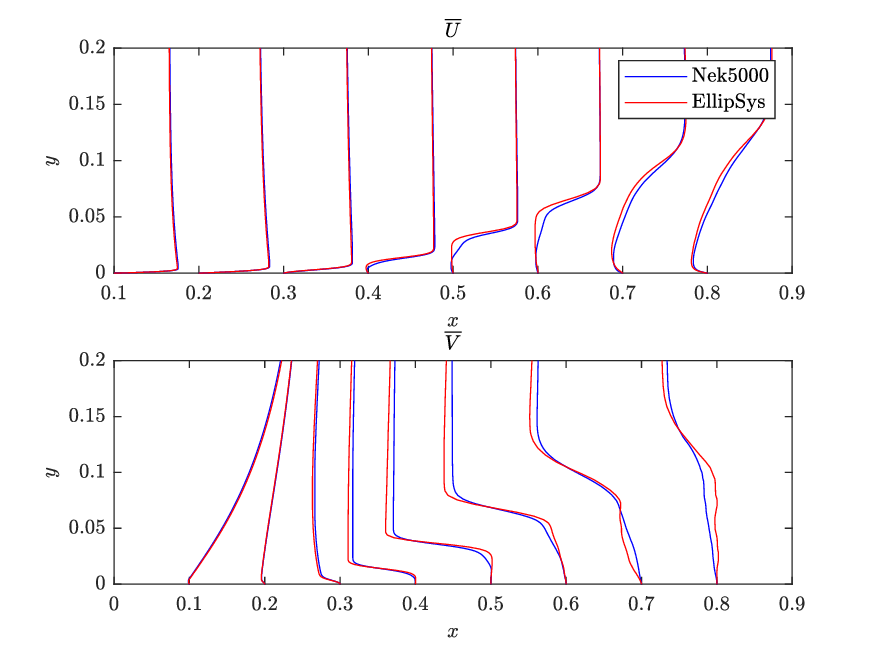}
	\vspace{-1cm}
	\caption{Comparison of the normal profiles of streamwise and normal velocities on the pressure side. A scale of 0.1 corresponds to a velocity of 1.4 in the plot of $\overline{U}$ and 0.2 in the one of $\overline{V}$.}
    \label{fig:vel_prof_PS_comparison}
\end{figure}

Table~\ref{tab:xb1_xb2_xtr} displays the comparison of mean locations of separation ($x_{b_1}$), reattachment ($x_{b_2}$), and transition ($x_{tr}$). The LSB edge is defined as $(x,y_b)$ fulfilling $\int_{0}^{y_b} \langle U\rangle_{z, t}(x,\xi) \; d\xi=0$ \citep{avanci2019}, i.e., $y_b$ is the zero streamwise mass-flux line. The transition location is assumed to be the streamwise locus of maximum boundary layer shape factor $H=\delta^*/\theta$, where $\delta^*$ is the displacement thickness and $\theta$ is the momentum thickness \citep{jaroslawski2023}. On the suction side, separation starts at $x_{b_1}=0.374-0.381$, and the percentual difference between \ellipsys{} and \nek{} $\Delta=-1.837\%$ is small. The same is valid for the reattachment point $x_{b_2}=0.801-0.802$, whose difference $\Delta=0.125\%$ is even lower. A larger discrepancy between the methods occurs for the transition location, which is 6.811\% more downstream in \ellipsys{} than in \nek{}. This is likely related to a lower averaging time in \nek{}. On the pressure side, the separation and reattachment locations are in close agreement. Notice that the flow remains separated until the trailing edge. The transition location for \ellipsys{} was estimated as the location of maximum  $u'_{RMS}$ since the peak in $H$ was not clear. This yielded $x=0.8^\dagger$, which is close to the transition location in \nek{} ($x_{tr}=0.8114$), obtained from the location of maximum $H$.

\begin{table}
  \begin{center}
  \begin{tabular}{lcccccc}
                    & \multicolumn{3}{c}{Suction side} 
                    & \multicolumn{3}{c}{Pressure side}\\
      Code          & $x_{b_1}$ & $x_{b_2}$ & $x_{tr}$ 
                    & $x_{b_1}$ & $x_{b_2}$ & $x_{tr}$\\[3pt]
      \ellipsys{}   & 0.374     & 0.802     & 0.602    
                    & 0.317     & 0.995     & $0.8^\dagger$\\
      \nek          & 0.381     & 0.801     & 0.561    
                    & 0.319     & 1.000     & 0.8114\\
      $\Delta$ (\%) & -1.837    & 0.125     & 6.811    
                    & -0.631    & 0.503     & --
  \end{tabular}
  \caption{Mean laminar separation bubble and transition locations.}
  \label{tab:xb1_xb2_xtr}
  \end{center}
\end{table}

\subsection{Perturbation evolution}

\begin{figure}
    \centering
    \begin{subfigure}[t]{1\textwidth}
        \includegraphics[width=1\textwidth]
        {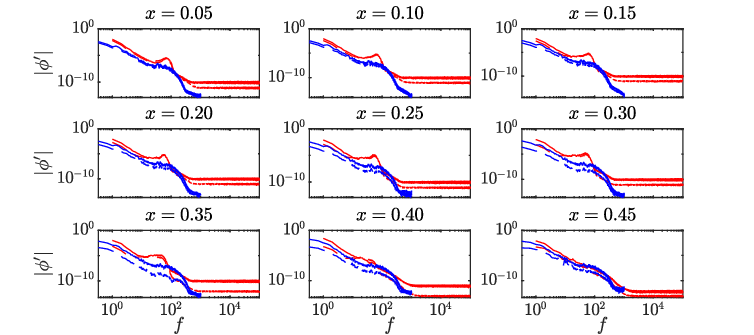}
        \label{fig:PSD_EllipSys_L1_o135n_bc0_SS_UV_1_final}
    \end{subfigure}
    
    \begin{subfigure}[t]{1\textwidth}
        \includegraphics[width=1\linewidth]
        {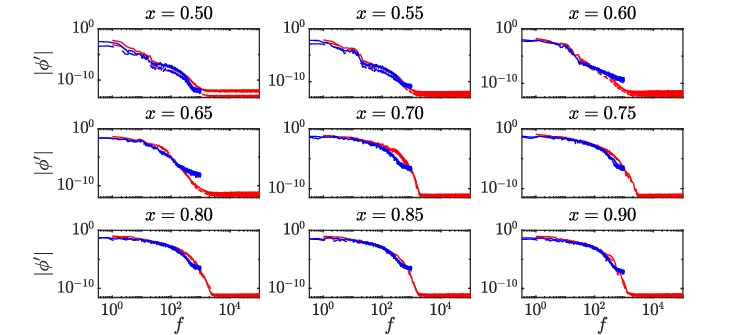}
        \label{fig:PSD_EllipSys_L1_o135n_bc0_SS_UV_2_final}
    \end{subfigure}%
    \caption{Amplitude spectra of velocity perturbations on the suction side at several streamwise stations. \protect\redline{} and \protect\dashedredline{} indicate streamwise and normal velocity perturbations, respectively, obtained with \nek{}. \protect\blueline{} and \protect\dashedblueline{} are the corresponding perturbations obtained with \ellipsys{}.}
    \label{fig:PSE_Nek_EllipSys_SS}
\end{figure}

\begin{figure}
    \centering
    \begin{subfigure}[t]{1\textwidth}
        \includegraphics[width=1\textwidth]
        {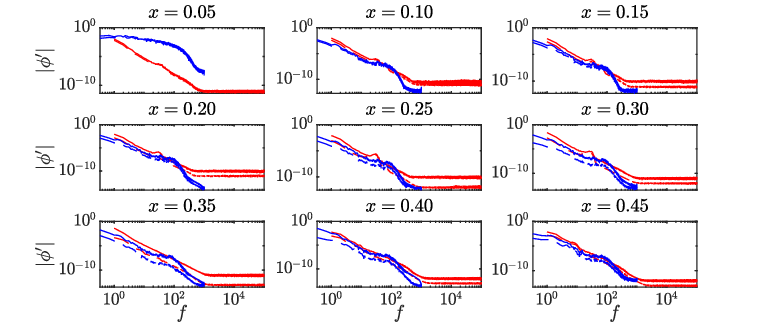}
        \label{fig:PSD_EllipSys_L1_o135n_bc0_PS_UV_1_final}
    \end{subfigure}
    
    \begin{subfigure}[t]{1\textwidth}
        \includegraphics[width=1\linewidth]
        {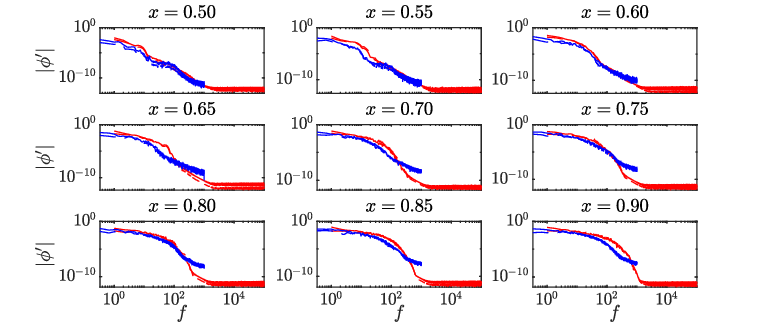}
        \label{fig:PSD_EllipSys_L1_o135n_bc0_PS_UV_2_final}
    \end{subfigure}%
    \caption{Amplitude spectra of velocity perturbations on the pressure side at several streamwise stations. \protect\redline{} and \protect\dashedredline{} indicate streamwise and normal velocity perturbations, respectively, obtained with \nek{}. \protect\blueline{} and \protect\dashedblueline{} are the corresponding perturbations obtained with \ellipsys{}.}
    \label{fig:PSE_Nek_EllipSys_PS}
\end{figure}

Figure~\ref{fig:PSE_Nek_EllipSys_SS} presents a comparison between the boundary layer spectra on the suction side obtained at a wall-normal height of $y=5\times10^{-3}$ with \nek{} (red line) and \ellipsys{} (blue line). Streamwise (solid line) and normal (dashed line) velocity perturbations are considered. In the attached laminar boundary layer in the region $x=0.05-0.35$, the spectra present differences around $f=28$, likely corresponding to Tollmien-Schlichting (TS) waves, in which \nek{} displays higher perturbation amplitude. Notice that these perturbations are very weak and reach a maximum amplitude $\mathcal{O}(10^{-6})$, being damped by numerical dissipation in \ellipsys{}. Since \nek{} is a pseudo-spectral code, it exhibits nearly zero dissipation.
There is good agreement between the codes in the separated flow region ($x=0.4-0.8$) and the reattached turbulent boundary layer ($x>0.8$). In particular, the energy peak relative to Kelvin-Helmholtz (KH) modes is well captured by both simulations and occurs around $f=10$ at $x=0.55$, for example. Regarding the results on the pressure side, shown in Fig.~\ref{fig:PSE_Nek_EllipSys_PS}, the initial perturbation level ($x=0.05$), corresponding to numerical noise, is higher in \ellipsys{}, as expected. Further downstream, the perturbation amplitude tends to the same values as boundary-layer instabilities develop. There is an overall good agreement between the results, apart from the peak at $f=28$ in the attached boundary layer, which is not predicted by \ellipsys{}, similarly to the suction side, due to the higher dissipation in this solver.

\subsection{Linear stability analysis}

\begin{figure}
    \centering
    \begin{subfigure}[t]{0.5\textwidth}
        \includegraphics[width=1\textwidth]
        {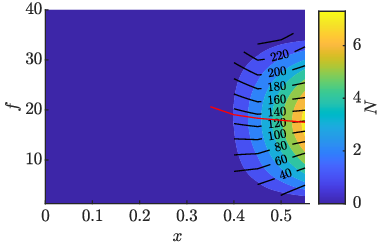}
        \caption{\ellipsys{}.}
        \label{fig:neutral_curve_spat_PSE_new_scl_factors_ref9_bis}
    \end{subfigure}%
    \begin{subfigure}[t]{0.5\textwidth}
        \includegraphics[width=1\linewidth]
        {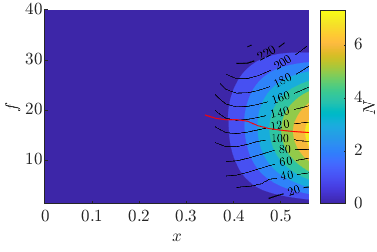}
        \caption{\nek.}
        \label{fig:neutral_curve_spat_PSE1_ref7}
    \end{subfigure}%
    \caption{Contours of $N$ factor from PSE. Isolines of the real part of the streamwise wavenumber are show in black. Locus of maximum amplification is highlighted with a red line.}
    \label{fig:neutral_curves}
\end{figure}

The parabolized stability equations \citep{bertolotti1991,bertolotti1992,herbert1997} were employed to compute the evolution of convective modes \citep{huerre1990} on the suction side of the blade. These modes consist mainly of TS and KH instabilities. Computations were carried out with the NOLOT PSE code \citep{hanifi1994}. The spanwise- and time-averaged flow was used as base flow for the stability analyses, and only two-dimensional perturbations ($\beta=0$) were considered as they are the most unstable ones \citep{squire1933}. Figure~\ref{fig:neutral_curves} shows the $N$ factor, which is a measure of the integrated growth of perturbations, as a function of the temporal frequency and streamwise location. The two neutral curves display good agreement, and the $N$ factor reaches 7.3 for $f=15$ at the most downstream station analyzed.

\begin{figure}
    \centering
    \begin{subfigure}[t]{0.5\textwidth}
        \includegraphics[width=1\textwidth]
        {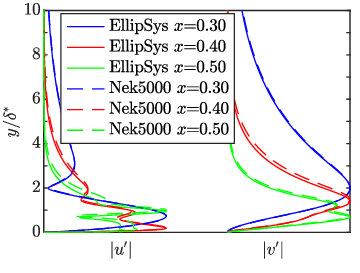}
        \caption{Wall-normal profiles of PSE modes.}
        \label{fig:eigenfunctions_monomode1_EllipSys_L1_o135n_bc0_f_3}
    \end{subfigure}%
    \begin{subfigure}[t]{0.5\textwidth}
        \includegraphics[width=1\linewidth]
        {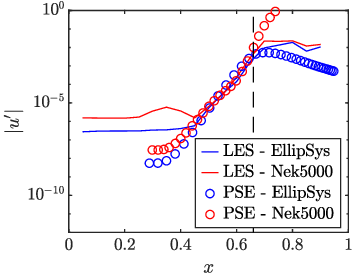}
        \caption{Streamwise evolution of perturbation.}
        \label{fig:uprime_versus_x}
    \end{subfigure}%
    \caption{Comparison between PSE and LES computations from \nek{} and \ellipsys{} for the mode with the highest $N$ factor ($f=15$, $\beta=0$) on the suction side. The vertical line in panel (b) indicates the streamwise location where the perturbation reaches $\mathcal{O}(10^{-2})$.}
    \label{fig:profiles_growth}
\end{figure}

Figure~\ref{fig:eigenfunctions_monomode1_EllipSys_L1_o135n_bc0_f_3} presents a comparison between the wall-normal profiles of the $f=15$, $\beta=0$ mode (streamwise and wall-normal velocity perturbations) at several streamwise stations obtained with the mean flows from \ellipsys{} and \nek{}. There is close agreement between the two sets of results for all streamwise positions. It is clear that upstream of separation ($x=0.3$), the mode corresponds to a TS wave. This changes at the front part of the LSB ($x=0.4$), where the profile presents a near-wall peak from a TS contribution and a second peak at the inflection point location. The latter peak is characteristic of a KH mechanism, which is inviscid and presents maximum production of perturbation kinetic energy at the inflection point location \citep{dovgal1994}. The TS and KH mechanisms coexist in this region, where the separated shear layer is thin and viscous effects are relevant \citep{rist_vki,diwan2009}. Further downstream, at $x=0.5$, the separated shear layer moves sufficiently far away from the wall so that the inviscid instability becomes the dominant driver of the perturbation kinetic energy production.
One can observe that the amplitude at the inflection point location becomes larger than that of the near-wall peak.

The streamwise evolution of the  $f=15$, $\beta=0$ mode is computed with LES and PSE. The results are presented in Fig.~\ref{fig:uprime_versus_x}. Despite the higher amplitude in the attached boundary layer in the \nek{} results, there is a close agreement for the growth rates computed with this solver and \ellipsys{} in the region where the mode displays the highest growth. This region corresponds to the LSB, where the KH mechanism displays the highest amplification. The agreement remains good along the streamwise extent where non-linear effects are small, which corresponds to the area with perturbation amplitude lower than $\mathcal{O}(10^{-2})$. The location where this occurs is marked with a vertical dashed line. Downstream of that, the mode saturates in the LES simulations, and the mean flow is not representative anymore of the base state where the perturbation grows.

\begin{figure}
	\centering
	\includegraphics[width=\linewidth]{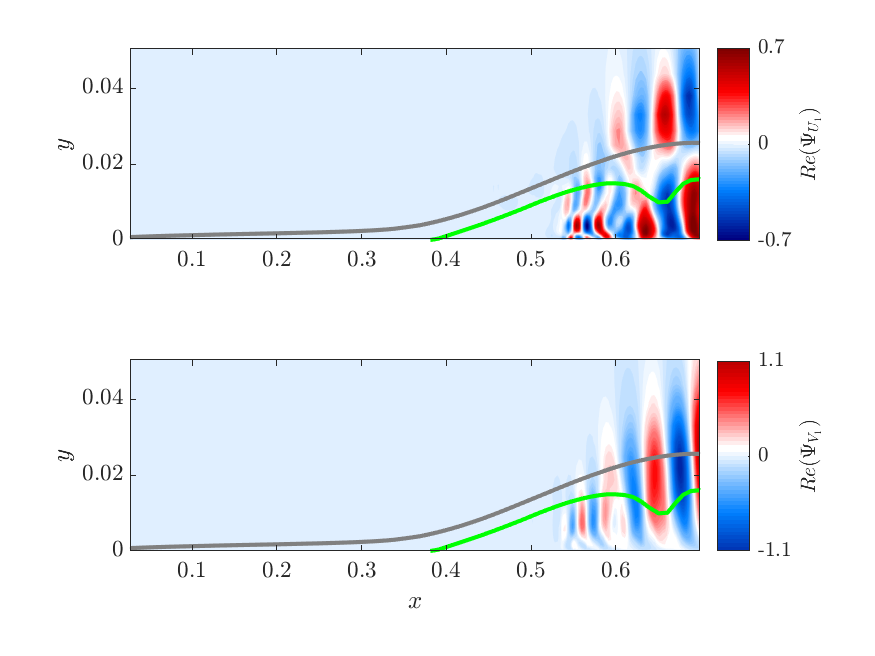}
	\vspace{-1cm}
	\caption{Real part of the first SPOD mode for streamwise and normal velocity perturbations for $f=15$, $\beta=0$. The amplitude of the mode corresponds to $\lambda_1$, i.e., the eigenvalue relative to the first mode. The gray line indicates the displacement thickness ($\delta^*$), and the green line denotes the LSB edge ($y_b$).}
    \label{fig:SPOD_L1_o135n_SS_f012_b0_p1_x0_07_yx_typex3_2}
\end{figure}

The spatial structure of the $f=15$, $\beta=0$ mode in the LES simulations is investigated with spectral proper orthogonal decomposition (SPOD) \citep{towne2018,schmidt2020}. This method extracts the most energetic coherent structures, ordering them from the highest (mode 1) to the lowest energy. The analysis domain consists of a region extending from the leading edge to 70\% chord on the suction side. 2,000 snapshots spaced with a time step $\Delta t=5\times10^{-3}$ are divided into ten blocks with $80\%$ overlap. The real part of the first SPOD mode for $f=15$, $\beta=0$ is portrayed in Fig.~\ref{fig:SPOD_L1_o135n_SS_f012_b0_p1_x0_07_yx_typex3_2}, where the streamwise ($\Psi_{U_1}$) and wall-normal ($\Psi_{V_1}$) velocity components are presented. The modes agree with previous investigations of the KH instability in LSBs, where $y=\delta^*$ lies approximately at the location of the phase inversion of $\Psi_{U_1}$ and maximum amplitude of $\Psi_{V_1}$ \citep{fava2023a,fava2023b,fava2023c}. 

\section{Long-term evolution of the normal force}\label{sec:long_term}

\begin{figure}
    \centering
    \includegraphics[width=0.9\textwidth]
        {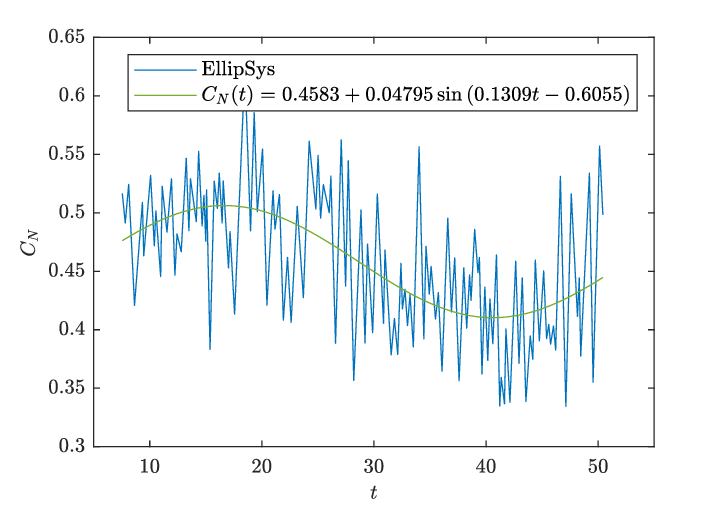}
    \caption{Evolution of the normal force coefficient. SIN is the function $C_{N}(t)=0.4583+0.04795 \sin\left(0.1309 t - 0.6055\right)$.}
    \label{fig:evolution_forces_EllipSys_2}
\end{figure}

Once validated, \ellipsys{} is employed to compute the flow development in the long term. The total time of the simulation is $T=50 c/U_\infty$ (50 flowthroughs). In order to remove the transient from the initial condition, the first ten flowthroughs are discarded. The evolution of the normal force coefficient ($C_N$) as a function of time is presented in Fig.~\ref{fig:evolution_forces_EllipSys_2}. Short-time scales of the order of one flowthrough can be seen in the $C_N$ curve, associated with the shedding of coherent structures from the separated shear layers \citep{raju2008}. In addition, there is a slow modulation of the mean $C_N$, which may correspond to the phenomenon of low-frequency oscillations (LFOs). The mean variation of the $C_N$ approximately follows a sinusoidal law with angular and temporal frequencies $\omega_f=0.1309$ and $f_f=0.021$, respectively, with a corresponding period $T_f=48$. This frequency is about 720 times lower than the frequency of the most unstable KH mode.
Furthermore, this period corresponds to 7.7 rotations of the DTU 10-MW Reference Wind Turbine at rated rotor speed  \citep{bak2012}. The widely employed Strouhal number $St=(f^* c/U_\infty) \sin AoA$ \citep{zaman1987} is computed, yielding $St=0.0012$, which is in the order of magnitude of LFO in the literature.
LFO is attributed to a feedback mechanism between the viscous boundary layer and the inviscid flow, with the alternation between regions of stalled and unstalled flow on the suction side \citep{zaman1989,broeren1998,sandham2008}. The shift from short LSB to fully stalled flow occurs via bursting \citep{mccullough1951,gaster1967,aniffa2023}. Bursting has been associated with the occurrence of absolute instability \citep{almutairi2013,negi2018a}, which may happen here due to the reverse flow of  $-17\%$ (time average) on the suction side \citep{hammond1998,alam2000,rist2002,fasel2004,diwan2009,rodriguez2013}. The amplitude of variation of the $C_N$ in the current case corresponds to approximately $10.5\%$ of its mean value. This is lower than the $50\%$ found in some cases in the literature \citep{zaman1989,bernardini2016}. A possible explanation for that is that the airfoil is in a post-stall condition.

\begin{figure}
    \centering
    \includegraphics[width=0.9\textwidth]
        {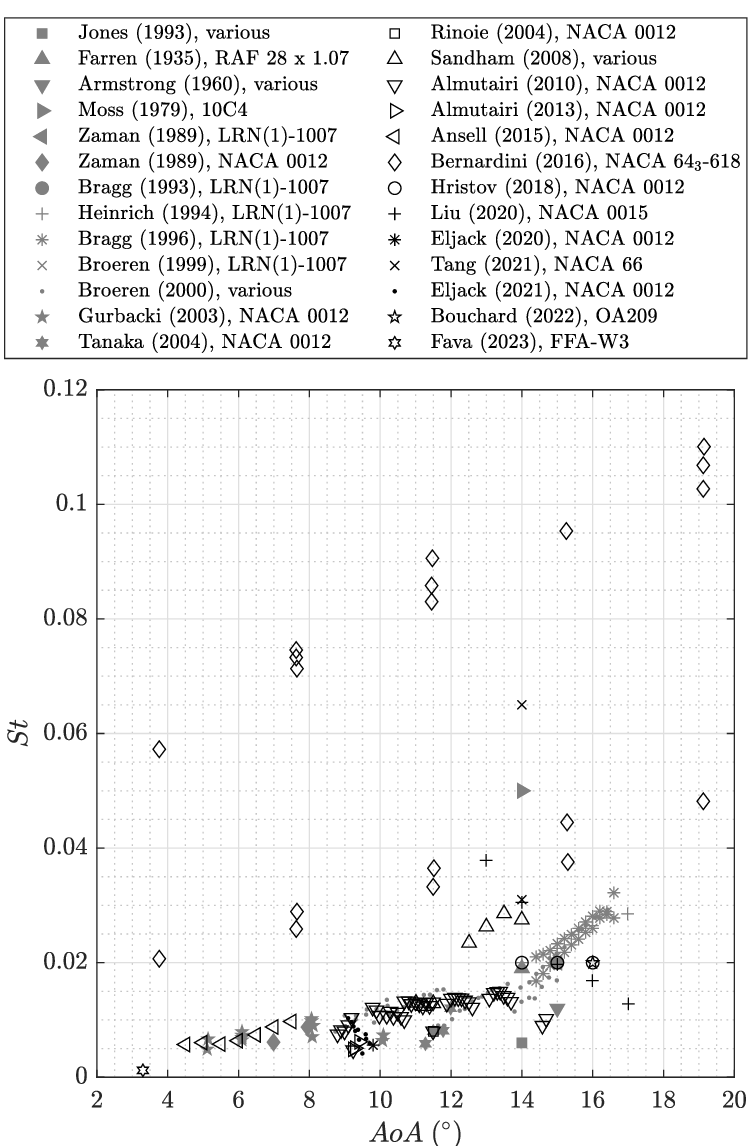}
    \caption{Compilation of $St=(f^* c/U_\infty) \sin AoA$ from the literature and current work (Fava (2023), FFA-W3).}
    \label{fig:strouhal_AoA}
\end{figure}

A compilation of $St$ as a function of the $AoA$ with results from the current work (Fava (2023), FFA-W3) and literature is presented in Fig.~\ref{fig:strouhal_AoA}. Most of the studies were performed for the NACA 0012 and LRN(1)-1007 airfoils, where a well-defined trend of increase in $St$ with $AoA$ exists. Nevertheless, other types of profiles, such as the NACA 64$_3$-618 airfoil and NACA 66 hydrofoil, display higher $St$ for the LFO, suggesting that the scaling with $St$ depends on the airfoil geometry. The current work is situated in the bottom left-hand corner of the diagram, with the lowest $AoA$ and $St$ among the results found in the literature. The reason for the appearance of such behavior at low $AoA$ is the thickness of $24\%$ of the current airfoil, which is twice that of the NACA 0012 profile. This leads to stall at relatively low incidence angles (maximum lift at $AoA=1.4^\circ$ according to XFOIL). Notice, however, that only the pressure side displays a trailing-edge stall in the current simulations. In contrast, the flow reattaches on the suction side, forming an LSB. The figure does not display the dependence of $St$ on the Reynolds number since this variation is small \citep{bragg1996}.

\section{Conclusions}\label{sec:conclusions}

The current work studies the transition and low-frequency oscillations (LFOs) phenomena on a wind turbine airfoil. The profile consists of a blend of $96\%$ of the FFA-W3-241 and $4\%$ of the FFA-W3-301 as employed at 68\% of the radius of the DTU 10-MW Reference Wind Turbine. Wall-resolved large eddy simulations (LES) of this blade section are conducted with the flow solvers \nek{} and \ellipsys{} with the chord Reynolds number of $1 \times 10^5$ and effective angle of attack of $3.1^\circ-3.3^\circ$.

The domain width of $10\%$ of the chord was proven enough to capture the flow physics compared to a domain twice wider. Furthermore, \ellipsys{} in LES mode was validated against \nek{} since the former solver is less computationally demanding. There was close agreement of the mean velocity fields, although small discrepancies were noted in the separated flow region. These minor differences were attributed to the averaging time, limited by \nek{}, and slight variations in the angle of attack. Regarding the perturbation evolution, \ellipsys{} also predicted close results to that from \nek{}, although a peak amplitude in the spectra, corresponding to Tollmien-Schlichting (TS) waves, was underpredicted by \ellipsys{}, probably due to higher numerical dissipation. Transition on the suction side was found to be caused by Kelvin-Helmholtz (KH) instability developing on the separated shear layer. The most amplified KH mode presented a frequency $f\approx15$, with its downstream evolution being closely predicted by the parabolized stability equations (PSE). Furthermore, both solvers indicated very close results.

In the following, \ellipsys{} was employed to compute the long-term evolution of the flow for about 50 flowthroughs. The normal force coefficient ($C_N$) evolution presented a slow oscillation with an amplitude of $10.5\%$ of the mean and a frequency $f=0.021$. This converts to $St=(f^* c/U_\infty) \sin AoA = 0.0012$ for an effective angle of attack of $3.3^\circ$.
This value of $St$ is in the range of LFO previously observed in airfoils. Nevertheless, from all the compiled results from the literature, the current $St$ is the lowest.
The oscillation period corresponds to 7.7 rotations of the DTU 10-MW Reference Wind Turbine at rated rotor speed. These oscillations have been attributed in the literature to the periodic stalling and unstalling of the suction side of the airfoil. The current analysis only noticed a full trailing edge stall on the pressure side, whereas the suction side presented a laminar separation bubble. The reverse flow on both sides was high enough to allow absolute instability, which could explain a periodic bursting of the LSB into fully stalled flow.


\bibliographystyle{jfm}
\bibliography{paper}

\end{document}